# Magnetically-Sensitive Valley Polarization Reversal and Revival of Defect-Localized Excitons in WSe$_2$-WS$_2$ Heterostructure

Taishen Li[1,#], Tao Yu[2,#], Xuefeng Cui[1], Kaixuan Zhang[1], Jianyi Liu[1], Qiushi Meng[1], Hongbing Cai[1,3], Nan Pan[1,3,*], Bing Wang[1,3], Zhenchao Dong[1] and Xiaoping Wang[1,3,*]

[1]*Hefei National Laboratory for Physical Sciences at the Microscale, University of Science and Technology of China, Hefei, Anhui 230026, P.R. China,*
[2]*Kavli Institute of NanoScience, Delft University of Technology, 2628 CJ Delft, The Netherlands*
[3]*Key Laboratory of Strongly-Coupled Quantum Matter Physics, Chinese Academy of Sciences, School of Physical Sciences, University of Science and Technology of China, Hefei, Anhui 230026, P. R. China.* [#]*These authors contributed equally to this work.*

[*]*email:* xpwang@ustc.edu.cn

*email:* npan@ustc.edu.cn

## Abstract

Manipulating and reserving the valley pseudospin of excitons is one core aim in the two-dimensional transition metal dichalcogenides (TMDs). However, due to the strong electron–hole exchange and spin-orbit coupling interactions, the exciton recombination lifetime is subject to picosecond timescale intrinsically, and the valley polarization is hardly modulated by a moderate magnetic field. It is fortunate that interlayer and defect-localized excitons promise to overcome these difficulties by suppressing the above interactions. Here we clearly reveal that the valley polarization can be reversed and revived in the defect-localized excitons with a microsecond lifetime in AB-stacked WSe$_2$-WS$_2$ heterobilayer. Specifically, for the interlayer defect-localized exciton, the valley polarization is reversed and can be efficiently enhanced by a weak out-of-plane magnetic field (< ±0.4 T). In sharp contrast, the valley polarization of the intralayer defect-localized exciton can revive after a fast decay process and follows the direction of the moderate out-of-plane magnetic field (< ±3 T). We explain the reversed valley polarization with highly magnetic sensitivity by the delocalization of defect-localized holes under a weak magnetic field and the revival of valley polarization by the valley Zeeman effect under a moderate magnetic field. Our results demonstrate that the valley pseudospin of defect-localized excitons can be efficiently modulated by the external magnetic field and enrich both the understanding and the technical approaches on manipulating the valley dynamics in TMDs and their heterostructure.

## Introduction

The electronic, optical and magnetic properties of the atomically thin transition metal dichalcogenides (TMDs) in the form of MX$_2$ (M=Mo, W; X=S, Se) have been studied intensively in the past few years[1-6]. For monolayer TMDs, benefiting from the strong spin-orbit interaction and broken inversion symmetry, the spin and valley degrees of freedom are coupled to each other, enabling a circularly optical access to the valley pseudospin by the valley-dependent optical selection rules[7]. Using the valley pseudospin coupled with the excitonic states, the TMDs have emerged as promising candidates for storing, encoding and proceeding information. However, owing to the strong electron–hole exchange interactions in monolayer TMDs, the valley polarization lifetime is intrinsically limited to picosecond[8]. Such short valley polarization lifetime seriously limits the development of valleytronic devices where a long valley lifetime is necessary to ensure the processing of the valley pseudospin before its dissipation and/or decoherence. Therefore, constructing architectures with higher valley polarization and longer valley lifetime holds the key for practical applications.

Fortunately, it has been proposed very recently that the TMDs' van der Waals (vdW) heterostructures, realized by either viscoelastic stamping or heteroepitaxial growth of two different monolayer TMDs, possess the unique advantages to circumvent these limitations[9-10]. Specifically, for type II band alignment, electrons and holes find their energy minimums in different layers[11] and the spatial separation significantly reduces the overlap of electron and hole wavefunctions and changes the excitonic properties [12]. Moreover, in this heterobilayer configuration, it allows to engineer the valley properties of the interlayer exciton by tuning a series of degrees of freedom, such as changing the geometric arrangement of bilayer stacking, selecting different TMDs as the components, imposing an external electric field and so on[13-15]. In parallel, it has also been demonstrated very recently that, defects in monolayer TMDs could be exceptionally desirable for various desired properties[16-19], and the excitons localized at which can behave as single-photon emitters[20-21], and/or possess a microsecond valley lifetime[22]. Thus, it is natural to ask: if a heterobilayer of TMDs was composed of the monolayers with certain defects, could these intralayer defects play any role in tuning the valley properties of the interlayer exciton. As far as we know, such kind of study on the role of defects in the heterostructures of TMDs has not yet been explored.

In the present work, we explored the magnetically-sensitive valley dynamics of the defect-localized excitons in the AB stacked $WSe_2$-$WS_2$ heterobilayer through both the steady-state and time-resolved photoluminescence (PL) measurements with circular polarization resolution. We have found that both the interlayer ($IX_{def}$) and intralayer ($X_{def}$) defect-localized excitons possess a microsecond-scale valley lifetime, exhibiting intriguing magnetically-sensitive responses in their valley dynamics. On one hand, the valley polarization of $IX_{def}$ is reversed and can be significantly enhanced by a weak out-of-plane magnetic field ($B_z < \pm 0.4$ T), presenting a sharp "Λ" pattern. On the other hand, the valley polarization of $X_{def}$ can revive after a fast relaxation process and follows the direction of a moderate out-of-plane magnetic field ($B_z < \pm 3$ T) showing an "X" pattern. We explained the magnetically-sensitive valley polarization reversal of the interlayer defect-localized exciton by the delocalization of defect-bound holes under a weak magnetic field, which then hopes to another layer to emit the oppositely polarized light. While the revival of valley polarization of the intralayer defect-localized exciton roots in the valley Zeeman effect under a moderate magnetic field.

## Results and discussion

The monolayer flakes of WSe$_2$ and WS$_2$ used in this study were grown respectively onto the SiO$_2$/Si substrates by conventional chemical vapor deposition (CVD) method, and then were stacked to form the heterobilayer in an AB alignment by a wet transfer method through polymethy methacrylate (PMMA)[23] using a homemade high-precision transfer platform (see method section). The optical microscope image of the heterobilayer is displayed in Fig. 1a, and the AB-stacking sequence has been confirmed by the result of second harmonic generation (SHG) measurement (see Fig. S1). Fig. 1b shows the PL spectra of the monolayer WS$_2$, WSe$_2$ and heterobilayer zones at the temperature of 10 K, respectively, with a 532 nm laser excitation and a spot size of ~1 μm. From these PL spectra, one can clearly identify that the peaks at 615 nm and 736 nm correspond to the intralayer excitons of monolayer WS$_2$ and WSe$_2$, respectively. Whereas the newly arising peak at 865 nm exclusively observable in the heterobilayer region is naturally attributed to the interlayer exciton, whose radiative recombination should involve an electron at the bottom of conduction band of the monolayer WS$_2$ and a hole on the top of valence band of the monolayer WSe$_2$[9, 24-25].

It is worth noting that the full width at half maximum (FWHM) of the PL (at 736 nm) from monolayer WSe$_2$ is about 65 meV at 10K, which is twice in magnitude larger than the one of the monolayer WS$_2$ (at 615 nm). Additionally, through the fitting result of the PL of the monolayer WSe$_2$ (Fig. 1c), one can clearly observe a mild peak at 706 nm except for the dominate one centered at 736 nm. According to literatures, the former corresponds to the free exciton (X) of monolayer WSe$_2$; whereas the latter is unlikely to be the well-studied trion since the observed energy separation is as large as ~65 meV, exceptionally larger than the reported values (only about 20-40 meV) for the trions in monolayer TMDs[26-28]. Therefore, it strongly suggests that the main peak at 736 nm comes from the defect-localized excitons (X$_{def}$) in the monolayer WSe$_2$. Basing the fact that an emission from defect-localized excitons can be saturated with intense pumping, the PL intensities for the WSe$_2$, WS$_2$, and heterobilayer as the functions of excitation power are plotted together to confirm their origins, and the results are shown in Fig. 1d (color dots). Such a power-dependence can be well fitted by the formula $I=P^\alpha$, where $I$ is the PL intensity, $P$ is the excited laser power, and α represents the fitting constant. From the fitting results, we can observe that the PL of WS$_2$ (at 615 nm) presents a linear relation with the excitation power (α~1.02), indicating the unsaturated property of the free excitons of monolayer WS$_2$. However, the PL intensities for the monolayer WSe$_2$ (at 736 nm) and the heterobilayer (at 865 nm) show similar sub-linear power dependence with the α of ~0.70 and 0.71, respectively. These features are exactly one of the signatures for defect-related PL emissions, which also suggest that these two emissions share the same defect origin: certain defect comes unambiguously from the monolayer WSe$_2$.

Considering that the defect-bound and free excitons have the different temperature dependent PL evolutions and could convert to each other following the Boltzmann distribution, we systematically studied the variable-temperature PL spectra of the WSe$_2$, WS$_2$, and heterobilayer (see Fig. S2 for the complete spectral survey). In detail, Fig. 1e and f plot the normalized variable-temperature PL spectra of both the interlayer exciton and the intralayer one in monolayer $WSe_2$, respectively. Unlike the usual temperature dependent monotonic redshift of an excitonic interband transition due to lattice thermal expansion induced bandgap narrowing, the spectra of these two excitons share a common blue shift "transition" temperature at 120~150 K (black dashed boxes in Figs. 1e and f). Figs. S4 and S5 show more details for the normalized PL together with the Gaussian fitting results at four representative temperatures extracted from Figs. 1e and f, respectively. From this fitting data and energy band diagram, the seeming "discrepancy" between (1) the very different temperature-dependent spectral evolutions and the (2) shared blueshift "transition" temperature for the two kinds of excitons can be well interpreted, and the detailed analysis is given in the supplementary information. Here, we only give a summarized physical interpretation. For the interlayer exciton (Fig 1e), the small red shift of about 20 meV is simply attributed to the band gap narrowing due to thermal expansion of the lattice as temperature increases (as shown in Figs. S4 and S6a). However, this is obviously not the case for the intralayer exciton (Fig. 1f), of which an exceptionally large redshift of ~120 meV is clearly seen below the "transition" temperature. In fact, as can be seen from the fitting results of Fig. S5, in addition to the bandgap narrowing induced redshift of about 20 meV (as shown in Fig. S6b), there is another and much more important process responsible for the large redshift: the dominant component of PL from 10 to150 K transforms from the shallow defect-localized exciton ($X_{def}$) to the deeper defect-localized exciton ($X_D$) due to the different thermal activation energies. As labeled in the band alignment in Fig. S3, the defect level is shallow-acceptor-like, exactly locating above the valence band maximum of the $WSe_2$[17]. The interlayer exciton is also bound by this defect (named as $IX_{def}$) at the low temperatures, where the radiative recombination actually involves an electron from the $WS_2$ conduction band minimum and a hole bound by the $WSe_2$ shallow-acceptor-like defect. Therefore, it is natural to observe the common blueshift "transition" behavior at almost the same temperature for $IX_{def}$ and $X_{def}$ (Figs. 1e and f), since the shallow-acceptor-bound holes could be thermally activated and excited back to the $WSe_2$ valence band around 120~150 K. Finally, we can clearly show the schematic of exciton configuration in real space of the AB-stacked $WSe_2$-$WS_2$ heterostructure in figure 1g.

In order to reveal the transition process of the defect related excitons, we further explore the time-resolved PL under different temperatures using a pulsed laser at the excitation wavelength of 670 nm with a spot size of ~1 μm. The experimental time-resolved PL spectra of monolayer $WSe_2$ and $WSe_2$-$WS_2$ heterobilayer are shown in Fig. 2a and b, respectively, from which we can clearly observe that the dynamics of the PL includes the fast and slow processes. The slow process is significantly suppressed as the increasing temperature. Therefore, it is necessary to use the biexponential formula to fit the exciton dynamics of the fast and slow process, in the form of[29]

$$I(t) = A_1 e^{\frac{-t}{\tau_1}} + \frac{B}{t+R} + A_2 e^{\frac{-t}{\tau_2}}, \tag{1}$$

where $\tau_1$ and $\tau_2$ are the lifetimes of the fast and slow process, respectively. $A_1$, $A_2$, $B$ and $R$ are fitting parameters related to the initial population and the rate constants. Fig. 2c and d show the typical fitting results for time-resolved PL of monolayer $WSe_2$ and $WSe_2$-$WS_2$ heterobilayer measured at 5K, respectively. We can clearly observe that the time-resolved PL can be well fitted by the double exponential formula (Eqn. 1). The fitting paramaters of the initial population ($A_1$ and $A_2$) corresponding to spectra of monolayer $WSe_2$ and $WSe_2$-$WS_2$ heterobilayer are shown in Figs. S8a and c, respectively. Through carefully comparing the variable-temperature steady (Figs. S2, 4 and 5) and time-resolved (Figs. 2a, b, and S8) PL data, one can clearly see that the initial population of the slow processes ($A_2$ in Fig. S8a and c) and the corresponding steady PL intensity of the defect-localized excitons ($X_{def}$ and $IX_{def}$) monotonously and rapidly decrease, and even disappear above 150 K. However, the initial population of the fast processes ($A_1$ in Figs. S8a and c) and the corresponding PL intensity of the free excitons (X and IX) only drop slightly. Therefore, it strongly indicates that the fast ($A_1$) and the slow ($A_2$) decay processes stem from the transitions of the free excitons (X and IX) and the defect-localized ones ($X_{def}$ and $IX_{def}$), respectively. In addition, the fitting results of the lifetimes (Figs. S8b and d) reveal that, the lifetime of the fast decay ($\tau_1$ in Fig. S8d) for the interlayer transition is about a few nanoseconds, which is nicely consistent with the previously reported lifetimes for IX[30], Whereas both lifetimes of the slow decay ($\tau_2$ Figs. S8b and d) are up to microsecond timescale, which are exceptionally longer than all of the reported values for free excitons, charged excitons (trions), and biexcitons in monolayer TMDs[31]. and therefore should correspond exclusively to the defect-localized excitons. Actually, K. L. Silverman *et al.* have reported very recently that, the defect-localized excitons in monolayer $WSe_2$ could have an ultralong lifetime (~300ns) [22], qualitatively well consistent with the observations here. It is worth noting that the increase in the lifetime of the slow process with the increase of temperature might be the unique properties of the defect-bound excitons. This is in contrast to the previously reported dark state-brightness caused by the splitting of the conduction band in $WSe_2$[32]. As a confirmation of our understanding, we introduced a large number of defects into monolayer $WSe_2$ sample by high-energy ray irradiation and tested the time-resolved PL of the irradiated sample at 10K (Fig. S7d). Undoubtedly, the irradiated monolayer $WSe_2$ sample possesses an ultralong lifetime (~1000ns), which is similar to Fig. 2a and b. With these understanding, the schematic of exciton transition process in the $WSe_2$-$WS_2$ heterostructure can be shown as Fig. 2e.

As we know, because the valley polarization is one core figure of merit in the valley dynamics[33-34], efficient modulation of the valley polarization has great significance. The valley or PL polarization is defined as

$$P_{\pm} = \frac{I(\sigma\pm) - I(\sigma\mp)}{I(\sigma\pm) + I(\sigma\mp)}, \tag{2}$$

where $I(\sigma\pm)$ presents the intensity of σ± circularly polarized PL detection. In order to evaluate the magnetic field response of valley polarization of the defect-localized excitons, we first measured the steady-state PL spectra in detection with σ- and σ+ circular polarization as a function of out-of-plane magnetic field ($B_z$) under σ+ excitation as shown in Figs. 3a and b (the spectra under σ- excitation shown in S9a and b). Comparing the circularly polarized PL spectra (Figs. 3a and b, Figs. S9a and b), it can be remarkably observed that although the attenuation dynamics of interlayer ($IX_{def}$) and intralayer ($X_{def}$) defect-localized excitons have many similarities (Figs. 2a and b), their magnetic response of valley polarization are quite different. The PL intensity of $IX_{def}$ (at 865nm) is sensitive to the magnitude but independent of the direction of the magnetic field ($B_z$). In detail, the PL detection with opposite circular polarization to the pumped one (shown in Figs. 3b and S9b) is significantly enhanced within ±0.4T and then becomes saturated with the further increase of the magnetic field. Nevertheless, the PL detection with the same circular polarization to the pumped one (shown in Figs. 3a and S9a) rapidly declines within ±0.4T, but becomes slower with larger magnetic field. Such a small magnetic field can have significant influence on the PL polarization is beyond the conventional valley Zeeman effect. On the contrary, the circularly polarized PL of $X_{def}$ (at 736nm) is dominated by the direction of $B_z$. As shown in Figs. 3a and b, the circularly polarized PL of the $X_{def}$ with σ+ excitation changes monotonously but reversely with the increase of $B_z$.

According to the definition of valley polarization (Eqn. 2), we can quantitatively calculate the valley polarization of the defect-localized excitons. Fig. 3c shows the calculated valley polarization of $IX_{def}$ as a function of magnetic field and wavelength with σ+ excitation (the one with σ- excitation shown in S9c). What makes this especially remarkable is the dark line around 0T in Figs. 3c and S9c, indicating that the polarization of $IX_{def}$ possesses a *negative* value (even in the absence of the magnetic field) and can be significantly enhanced by a weak magnetic field (< ±0.4 T). Fig. 3e shows the averaged valley polarization of $IX_{def}$ around central wavelength (865 nm) extracted from Figs. 3c and S9c. We can clearly observe that the steady-state valley polarization of $IX_{def}$ presents a "Λ"-like dependence on the weak magnetic field. We calculate the slope of the curve in Fig. 3e to be about 0.15/KGs. Such a high magnetic sensitivity indicates that the $IX_{def}$ possesses a unique (external field regulation) advantage in the application of the valley electronics device. In sharp contrast to $IX_{def}$, The valley polarization of $X_{def}$ (Figs. 3d and S9d) is obviously dominated by the direction of $B_z$ (< ±3 T). From the averaged results (Fig. 3f) extracted from Figs. 3d and S9d, the $B_z$ tuning dependent PL polarization presents an "X"-like tendency, which is similar to the research of free excitons[35]. The experimental valley polarization results reveal that the $IX_{def}$ and $X_{def}$ have an obviously different magnetic response and the defect-localized excitons possess some unique and unrevealed transition dynamics and merits comparing with free excitons.

In order to understand above steady-state results (Fig. 3), we need to figure out the transition dynamics of the defect-localized excitons. We perform the time and circular polarization resolved measurements for both of the interlayer and intralayer defect-localized excitons under different out-of-plane magnetic fields ($B_z$) using 670 nm pulse laser with repetition of 500 KHz. Fig. 4a shows the σ+ (red cure) and σ- (black cure) detected time-resolved PL of interlayer excitons ($IX_{def}$ and IX) at selected $B_z$ magnitudes with σ+ excitation respectively (the one with σ- excitation is referred to Fig. S10a). Consistent with the steady state PL of $IX_{def}$ in Fig. 3a and b, the slow decay process of σ- detected time-resolved PL is stronger than that of σ+ detected time-resolved PL under σ+ excitation after applying a weak $B_z$. However, the fast process is almost unaffected by the weak magnetic field. According to the definition of valley polarization in Eqn. (2), we calculate the time-resolved valley polarization for interlayer excitons with σ+ excitation at selected weak $B_z$, as shown in Fig. 4b (the one with σ- excitation is shown in Fig. S10b). Interestingly, it reveals that in the absence of $B_z$, the interlayer excitons have possessed a negative valley polarization at the beginning of the radiative recombination, which quickly decay to nearly zero within the next few nanoseconds. Therefore, there is a small negative valley polarization (-10%) observed in the steady-state PL polarization (Fig. 3e). When adding a weak $B_z$ within 0.4 T, we find that the slow decay process for the valley polarization is significantly enhanced. We fit the fast and slow processes in Fig. 4b and S10b with the bi-exponential decay

$$P(t) = C_1 e^{\frac{-t}{\tau_3}} + C_2 e^{\frac{-t}{\tau_4}}, \tag{3}$$

where $\tau_3$ and $\tau_4$ are the valley lifetimes of the fast and slow decay, and $C_1$ and $C_2$ are the initial polarization related parameters of these processes, respectively. Fig. 4c show the fitting results of $C_1+C_2$ (purple dots) and $C_2$ (orange dots) with σ+ excitation (the one with σ- excitation is plotted in Fig. S10e). We can observe that $C_2$ shows a "Λ"-like tendency with the increase of $B_z$, but $C_1+C_2$ is nearly unchanged with $B_z$. According to Fig. 2b, we know that the fast process is originated from free interlayer excitons (IX), but the slow process roots in defect-localized interlayer excitons ($IX_{def}$). Therefore, the magnetic response of $C_1$ is consistent with that of valley polarization of $IX_{def}$. Meanwhile, the slow decay lifetime ($\tau_4$) shown in Fig. S10e becomes longer (even up to 3000ns at 0.4T) showing a "V"-like pattern with the increase of $B_z$, however, the fast decay lifetime ($\tau_3$) shown in Fig. S10d is kept around 10 ns with a little fluctuation. It indicates that the valley pseudospin of $IX_{def}$ is more sensitive to the magnetic field ($B_z$) in comparison to the one of IX.

Recently, W. Gao et al. observed the microsecond polarization lifetime of interlayer exciton enhanced by the magnetic field in $MoSe_2$-$WSe_2$ AA-stacked heterobilayer[29]. Considering it was widely reported in monolayer $WSe_2$ that the dark exciton can emerge with energy lower than the bright one and can be brightened by a large inplane magnetic field[36]. It is reasonable to contribute the magnetic field enhanced polarization lifetime of interlayer excitons to the dark excitons in the monolayer $WSe_2$ at larger magnetic field (> 1 T). However, in this work, we find that the valley polarization of the interlayer defect-localized exciton possesses an unique magnetically-sensitive and polarization-reversal property, which is rarely observed up to now. Therefore, we mainly focus on the phenomena at the weak regime of an out-of-plane magnetic field ($B_z$ < 0.4 T). Also note that the reversed (or negative) PL polarization of the interlayer exciton in AB-stacked TMD heterobilayers was only theoretically predicted by W. Yao et al.[37] and was observed experimentally in AA stacking $WSe_2$-$MoSe_2$ heterostructure by W. H. Chang et al. recently[13]. Obviously, more in-depth experimental studies on this topic are urgently called for in the future. Very recently, X. Xu et al. observed a positive PL polarization of the interlayer moiré-trapped excitons in nearly AB-stacked $MoSe_2$-$WSe_2$ heterobilayer whereas a negative one in nearly AA-stacked sample[38]. It is exceptionally exciting to prospect that, in addition to the interlayer free excitons, the interlayer excitons trapped by certain potential wells could have more fascinating and even completely new valley properties. Besides the moiré-trapped interlayer exciton, the defect-bound interlayer exciton is indeed another member in this promising family, where the potential wells can naturally offer extra degrees of freedom to manipulate the valley pseudospin of the interlayer excitons. We can also explain our observed features for IX$_{def}$ using the optical dipole of the interlayer exciton with kinematic momentum **Q**, in which the electron and hole reside in the $\tau K$ and $-\tau K$ valleys in the WS$_2$ and WSe$_2$ layer ($\tau = +1$ or $-1$ represents the valley indexes). The optical dipole operator **D** is given by (derivation is referred to supplementary Eqns. S1~6)

$$\mathbf{D}_{-\tau K, \tau K}(\mathbf{Q}) \approx \langle 0 | \mathbf{D} | X^{(0)}_{-\tau K, \tau K}(\mathbf{Q}) \rangle + \frac{t_e}{E_I(\mathbf{Q}) - E_{WSe_2}} \langle 0 | \mathbf{D} | P_{\tau K}(\mathbf{q}) \rangle, \quad (4)$$
$$+ \frac{t_h}{E_I(\mathbf{Q}) - E_{WS_2}(\mathbf{q})} \langle 0 | \mathbf{D} | \Lambda_{-\tau K}(\mathbf{q}) \rangle$$

The first term of the right-hand side denotes the direct recombination of interlayer exciton $X^{(0)}_{-\tau K, \tau K}$ with the electron in the WS$_2$ monolayer and hole in the WSe$_2$ monolayer. The efficiency of the direct recombination is expected to be negligibly small due to the spatial separation of the electron and hole in different layers. The second term of the right-hand side comes from the interlayer hopping of the electron and the adjacent recombination of the intralayer exciton $P_{\tau K}$ in the WSe$_2$ layer, in which $t_e$ is the interlayer hopping energy of the electron, $E_I(\mathbf{Q})$ and $E_{WSe2}$ are the energy of the interlayer and WSe$_2$-intralayer excitons, respectively. This process does not cause the reversal of the PL polarization. While the third term of the right-hand side arises from the interlayer hopping of the hole and the adjacent recombination of the intralayer exciton $\Lambda_{-\tau K}$ in the WS$_2$ layer, in which $t_h$ is the interlayer hopping energy of the hole and $E_{WS2}$ is the energy of the WS$_2$-intralayer exciton. This process indeed causes the reversal of the PL polarization because the intralayer exciton $\Lambda_{-\tau K}$ in the WS$_2$ layer emits the PL with opposite polarization to that of $P_{\tau K}$ in the WSe$_2$ layer. Therefore, when the third process is dominant due to the more efficient interlayer hopping of the hole than the electron one, the PL polarization become reversed, as the observations in our experiments. Please note that although the optical dipole involves the hole or electron hopping process, these hopping processes are virtual or intermediate in quantum mechanics. Therefore, the emitted PL is still resonant to the interlayer exciton in energy, e.g., around 865 nm in wavelength here, but not the ones of the intralayer excitons.

As a special case to possibly understand the relative magnitudes of the hopping energy $t_e$ and $t_h$ to understand why the third process could be dominant, we discuss some limit cases as follows. In the commensurate heterobilayer, $t_e$ tends to be zero due to the interference effect during the electron hopping but $t_h$ is still large[39]. In reality, the lattice is not commensurate because the formation of interlayer exciton indicates a finite $t_e$, but the interlayer hopping of holes may still be more efficient than that of electrons as indicated by the first-principle calculation[37]. From this viewpoint, we attribute our measured negative PL polarization to the efficient interlayer hopping of hole that is also consistent with our interpretation on the magnetic-field sensitivity of the PL polarization, as we will discuss below. Thus, in Fig. 4d, consistent with Fig. 2b, the fast and slow processes can be understood according to the recombination of the intrinsic and defect-localized interlayer excitons due to the efficient interlayer hole hopping (more explanation is referred to supplementary S12).

From this viewpoint, any process that influences the interlayer hopping may influence the PL polarization in the heterobilayer, e.g., the interface quality or motion between the two layers. In particular, at the low temperature the weak localization of the hole arising from the quantum interference of different paths in the disordered $WSe_2$ layer by the W-defect inevitably influences the interlayer hopping of hole[17]. By a small out-of-plane magnetic field, the interference of different paths of the hole become destructive[40], and hence the weak-localization of the hole becomes less efficient and the interlayer hopping of the hole can play more important role in the optical dipole of the interlayer exciton [Eqn.~(4)]. This could be responsible for the enhancement of the recombination of the interlayer defect-localized exciton to emit PL with opposite polarization that is sensitive to a weak magnetic field in the slow process (shown in Figs. 4b and c). Nevertheless, the fast process is less influenced by the weak magnetic field because the motion of the relatively free hole cannot be efficiently influenced by a weak magnetic field.

As addressed in Figs. 3d and f, we have seen that the $B_z$-tunable PL polarization of $X_{def}$ presents an "X" pattern that is obviously different to the magnetic-field dependence of $IX_{def}$, suggesting their different transition dynamics. In order to figure out the microscopic process, we measure the time and circular polarization resolved PL of $X_{def}$ at 10K using the pulse laser (670 nm). Fig. 5a shows the σ+ (red) and σ- (black) detected time-resolved PL spectra of $X_{def}$ with different out-of-plane magnetic field ($B_z$) under the σ+ excitation (the one with σ- excitation is shown in Fig. S11a). The measurements reveal that the slow process of the PL detected in σ- polarization become stronger than the one in σ+ polarization when the $B_z$ increases towards the positive magnitude. Whereas, this relation becomes reversed when $B_z$ is reversed to the negative direction. Based on Fig. 5a, the time-resolved valley polarization of $X_{def}$ under σ+ excitation is plotted in Fig. 5b under different out-of-plane magnetic fields (the one with σ- excitation is shown in Fig. S11b). Again, the time-resolved PL polarization possesses the fast and slow processes. We see that the PL polarization of the slow process is even larger than that of the fast process at larger magnetic field, showing a revived phenomenon from fast to slow process. The revival of the PL polarization can be kept in a large magnitude (90%) without evident decay in a few microseconds. Moreover, the measurements show that the initial PL polarization of the fast process is less influenced by the magnetic field, but the PL polarization of the slow process becomes reversed when the direction of $B_z$ is from the positive to the negative direction. This suggests that the PL polarization of the slow process is related to the distribution of excitons determined by the magnetic field[35]. We note that the magnetic response of the slow process of PL polarization in Fig. 5b is consistent with that of the steady PL intensity in Fig. 3f. It further demonstrates that the fast process is attributed to the intralayer free exciton with lifetime being sub-nanosecond and no valley polarization reversal happens; the lifetime of the intralayer defect-localized exciton is extended to be hundreds of nanosecond, while its valley polarization lifetime is also in the order of microsecond[22].

The physics here can be understood from the valley Zeeman effect[41-42]. Fig. 5c summaries the schematics of transition dynamics of $X_{def}$ under zero, positive and negative out-of-plane magnetic field. The moderate out-of-plane magnetic field causes significant valley Zeeman splitting of the defect-localized intralayer excitons that influences their quasi-equilibrium distribution, i.e. they tend to populate at lower energy state. Their long lifetime in microsecond indicates that the quasi-equilibrium states could be achieved due to the magnon-phonon and magnon-magnon interactions. At the beginning of the transition, the initial polarization of the fast process is mainly determined by the intralayer free excitons, benefiting from their ultrafast transition rate. As the intralayer free excitons rapidly recombine, the PL polarization becomes dominant by the defect-localized excitons due to their extremely long life time, in which regime the strong magnetic field play an important role. When $B_z=0T$, without the time-reversal symmetry broken, the equilibrium distribution of the defect-localized exciton tends to be the same that does not contribute to the PL polarization in the slow process. A large magnetic field ($B_z \sim \pm 3T$) can cause a significant Zeeman splitting of these excitons in the two valleys, that then breaks the balance of these exciton populations in the two valleys by forming more low-energy exciton as indicated in middle and bottom panel of Fig. 5c. These low-energy exciton stores the valley polarization that can be even larger than the initial one with sign determined by the magnetic field direction. Therefore, we can observe the "X"-like magnetic response of steady PL polarization in Fig. 3f and the revival of time-resolved polarization in Fig. 5b.

## Conclusion

In summary, we have investigated the tunable PL polarization dynamics of the defect-localized excitons by out-of-plane magnetic field in the AB-stacked $WSe_2$-$WS_2$ heterostructure. Several interesting features related to the defect-localized excitons can be extracted from our experiment. Firstly, both the interlayer and intralayer defect-localized excitons possess microsecond radiative and valley polarization lifetimes. Secondly, the PL polarization of the interlayer defect-localized exciton is intrinsically negative and can be significantly enhanced in magnitude by the weak out-of-plane magnetic field ($< \pm 0.4$ T). Thirdly, the PL polarization of the intralayer defect-localized exciton is determined by the direction of out-of-plane magnetic field and even revives by a strong magnetic field (at $\pm 3$ T) in the time evolution: the valley polarization increases up to 90% within few nanoseconds and then keeps this value in about few microseconds without attenuation. We explain these defect-related abnormal phenomena in the delocalization of exciton by the weak magnetic field and valley Zeeman effect under stronger magnetic field. According to our work, defect-localized states of the TMDs heterostructure offer a broad and promising platform to extend the valley pseudospin lifetime and enrich the valley exciton physics.

## Methods

**Monolayer $WSe_2$ and $WS_2$ growth**
We firstly grow the monolayer $WSe_2$/$WS_2$ through chemical vapor deposition (CVD) method. In detail, at upstream high-purity Se/S powder (Alfa Aesar, 36208/10785) of 0.2 g was added into a small quartz boat that served as the evaporation precursor. At downstream high-purity $WO_3$ powder

(Alfa Aesar, 11828) of 10 mg was filled into another small quartz boat which was covered with the cleaned and face-down SiO$_2$/Si wafers (1.5× 1.5 cm$^2$ in size) as the substrates. The whole system (the chamber and pipes) was evacuated by a mechanical pump to a base vacuum of ~4.5 Pa, and then followed by the re-filling of ultrahigh-purity argon gas (~99.999%) with a constant flow of 120 sccm. Under the protective environment of this carrier gas, the two heating zones of the tube furnace were ramped to target temperatures of 350/320 (for the Se/S powder) and 825 °C (for the WO$_3$ powder), respectively. From then on, the argon gas flow was turn down and kept at 60 sccm, and the hydrogen gas flow was set at 10 sccm. The growth was typically lasted for 30 min before being finally stopped by shutting off the power of the furnace. The substrate was naturally cooled down to room temperature before the cutoff of the argon flow and system venting.

**WSe$_2$-WS$_2$ heterostructure transfer**

We firstly spin-coated polymethy methacrylate (PMMA) onto the growth WSe$_2$ substrate and dry under 180 °C environment lasting 15 min, and then gently soaked into the NaOH solution (2 mol/L) about 60 min for etching the SiO$_2$ away and separating the PMMA/WSe$_2$ from substrate. After separation, clean the PMMA/WSe$_2$ using the deionized water very slowly and carefully. We faxed the WS$_2$/substrate on a rotatable platform and smoothly tiled the PMMA/WSe$_2$ onto a quartz window. At last, we stack the WSe$_2$-WS$_2$ heterostructure in AB order through rotating the WS$_2$ platform and moving the PMMA/WSe$_2$ quartz window under our homemade rotatable and high-resolution transfer platform system. Finally, we have measured the Second Harmonic Generation (SHG) signal of the WSe$_2$-WS$_2$ heterostructure as shown in Fig. S1, and demonstrated the AB stacking order[43].

**Time-resolved PL measurement**

The attodry 1000 system was used to provide a low temperature (>4K) μ-PL spectrum testing platform. We employee the Clark MXR Impulse Laser to generate a 670 nm pulse laser with 500 KHz and pulse width less than 200ps and an electrical pulse as the start signal of the timer. We set up the circularity-resolved confocal μ-PL testing system and focus the 670 nm pulse laser using a 100x objective lens on the sample surface down to ~1 μm. The laser power on the sample was controlled at 20 μW for the excitation. The single photon detector (SPCM-AQRH-14) was used for collection and detection of the PL signals and switching the optical into electrical signals immediately, which was inputted into the timer as the end signal.

# References

1. Mak, K. F.; Xiao, D.; Shan, J., Light–valley interactions in 2D semiconductors. *Nature Photonics* **2018,** *12* (8), 451-460.
2. Yuan, H.; Wang, X.; Lian, B.; Zhang, H.; Fang, X.; Shen, B.; Xu, G.; Xu, Y.; Zhang, S. C.; Hwang, H. Y.; Cui, Y., Generation and electric control of spin-valley-coupled circular photogalvanic current in WSe2. *Nat Nanotechnol* **2014,** *9* (10), 851-7.
3. Yin, X.; Wang, Q.; Cao, L.; Tang, C. S.; Luo, X.; Zheng, Y.; Wong, L. M.; Wang, S. J.; Quek, S. Y.; Zhang, W.; Rusydi, A.; Wee, A. T. S., Tunable inverted gap in monolayer quasi-metallic MoS2 induced by strong charge-lattice coupling. *Nature Communications* **2017,** *8* (1).
4. Koo, W.-T.; Cha, J.-H.; Jung, J.-W.; Choi, S.-J.; Jang, J.-S.; Kim, D.-H.; Kim, I.-D., Few-Layered WS2 Nanoplates Confined in Co, N-Doped Hollow Carbon Nanocages: Abundant WS2 Edges for Highly Sensitive Gas Sensors. *Advanced Functional Materials* **2018,** *28* (36), 1802575.

5. Yang, H.; Kim, S. W.; Chhowalla, M.; Lee, Y. H., Structural and quantum-state phase transition in van der Waals layered materials. *Nature Physics* **2017,** *13* (10), 931-937.

6. Sangwan, V. K.; Lee, H. S.; Bergeron, H.; Balla, I.; Beck, M. E.; Chen, K. S.; Hersam, M. C., Multi-terminal memtransistors from polycrystalline monolayer molybdenum disulfide. *Nature* **2018,** *554* (7693), 500-504.

7. Xiao, D.; Liu, G. B.; Feng, W.; Xu, X.; Yao, W., Coupled spin and valley physics in monolayers of MoS2 and other group-VI dichalcogenides. *Phys Rev Lett* **2012,** *108* (19), 196802.

8. Yu, T.; Wu, M. W., Valley depolarization due to intervalley and intravalley electron-hole exchange interactions in monolayerMoS2. *Physical Review B* **2014,** *89* (20).

9. Rivera, P.; Yu, H.; Seyler, K. L.; Wilson, N. P.; Yao, W.; Xu, X., Interlayer valley excitons in heterobilayers of transition metal dichalcogenides. *Nat Nanotechnol* **2018**.

10. Rivera, P.; Seyler, K. L.; Yu, H.; Schaibley, J. R.; Yan, J.; Mandrus, D. G.; Yao, W.; Xu, X., Valley-polarized exciton dynamics in a 2D semiconductor heterostructure. *Science* **2016,** *351* (6274), 688-691.

11. Kozawa, D.; Carvalho, A.; Verzhbitskiy, I.; Giustiniano, F.; Miyauchi, Y.; Mouri, S.; Castro Neto, A. H.; Matsuda, K.; Eda, G., Evidence for Fast Interlayer Energy Transfer in MoSe2/WS2 Heterostructures. *Nano Letters* **2016,** *16* (7), 4087-4093.

12. Jin, C.; Kim, J.; Utama, M. I. B.; Regan, E. C.; Kleemann, H.; Cai, H.; Shen, Y.; Shinner, M. J.; Sengupta, A.; Watanabe, K.; Taniguchi, T.; Tongay, S.; Zettl, A.; Wang, F., Imaging of pure spin-valley diffusion current in $WS_2$-$WSe_2$ heterostructures. *Science* **2018,** *360* (6391), 893-896.

13. Hsu, W.-T.; Lu, L.-S.; Wu, P.-H.; Lee, M.-H.; Chen, P.-J.; Wu, P.-Y.; Chou, Y.-C.; Jeng, H.-T.; Li, L.-J.; Chu, M.-W.; Chang, W.-H., Negative circular polarization emissions from WSe2/MoSe2 commensurate heterobilayers. *Nature Communications* **2018,** *9* (1).

14. Hanbicki, A. T.; Chuang, H. J.; Rosenberger, M. R.; Hellberg, C. S.; Sivaram, S. V.; McCreary, K. M.; Mazin, II; Jonker, B. T., Double Indirect Interlayer Exciton in a MoSe2/WSe2 van der Waals Heterostructure. *ACS Nano* **2018,** *12* (5), 4719-4726.

15. Nayak, P. K.; Horbatenko, Y.; Ahn, S.; Kim, G.; Lee, J. U.; Ma, K. Y.; Jang, A. R.; Lim, H.; Kim, D.; Ryu, S.; Cheong, H.; Park, N.; Shin, H. S., Probing Evolution of Twist-Angle-Dependent Interlayer Excitons in MoSe2/WSe2 van der Waals Heterostructures. *ACS Nano* **2017,** *11* (4), 4041-4050.

16. Refaely-Abramson, S.; Qiu, D. Y.; Louie, S. G.; Neaton, J. B., Defect-Induced Modification of Low-Lying Excitons and Valley Selectivity in Monolayer Transition Metal Dichalcogenides. *Physical Review Letters* **2018,** *121* (16).

17. Zhang, S.; Wang, C. G.; Li, M. Y.; Huang, D.; Li, L. J.; Ji, W.; Wu, S., Defect Structure of Localized Excitons in a $WSe_{2}$ Monolayer. *Phys Rev Lett* **2017,** *119* (4), 046101.

18. Song, S. H.; Joo, M.-K.; Neumann, M.; Kim, H.; Lee, Y. H., Probing defect dynamics in monolayer MoS2 via noise nanospectroscopy. *Nature Communications* **2017,** *8* (1).

19. Wang, Q.; Zhang, Q.; Zhao, X.; Luo, X.; Wong, C. P. Y.; Wang, J.; Wan, D.; Venkatesan, T.; Pennycook, S. J.; Loh, K. P.; Eda, G.; Wee, A. T. S., Photoluminescence Upconversion by Defects in Hexagonal Boron Nitride. *Nano Lett* **2018,** *18* (11), 6898-6905.

20. He, Y. M.; Clark, G.; Schaibley, J. R.; He, Y.; Chen, M. C.; Wei, Y. J.; Ding, X.; Zhang, Q.; Yao, W.; Xu, X.; Lu, C. Y.; Pan, J. W., Single quantum emitters in monolayer semiconductors. *Nat Nanotechnol* **2015,** *10* (6), 497-502.

21. Koperski, M.; Nogajewski, K.; Arora, A.; Cherkez, V.; Mallet, P.; Veuillen, J. Y.; Marcus, J.; Kossacki, P.; Potemski, M., Single photon emitters in exfoliated WSe2 structures. *Nat Nanotechnol* **2015,**

*10* (6), 503-6.
22. Moody, G.; Tran, K.; Lu, X.; Autry, T.; Fraser, J. M.; Mirin, R. P.; Yang, L.; Li, X.; Silverman, K. L., Microsecond Valley Lifetime of Defect-Bound Excitons in Monolayer WSe2. *Physical Review Letters* **2018,** *121* (5).
23. Wang, K.; Huang, B.; Tian, M.; Ceballos, F.; Lin, M. W.; Mahjouri-Samani, M.; Boulesbaa, A.; Puretzky, A. A.; Rouleau, C. M.; Yoon, M.; Zhao, H.; Xiao, K.; Duscher, G.; Geohegan, D. B., Interlayer Coupling in Twisted WSe2/WS2 Bilayer Heterostructures Revealed by Optical Spectroscopy. *ACS Nano* **2016,** *10* (7), 6612-22.
24. Unuchek, D.; Ciarrocchi, A.; Avsar, A.; Watanabe, K.; Taniguchi, T.; Kis, A., Room-temperature electrical control of exciton flux in a van der Waals heterostructure. *Nature* **2018,** *560* (7718), 340-344.
25. Hong, X.; Kim, J.; Shi, S.-F.; Zhang, Y.; Jin, C.; Sun, Y.; Tongay, S.; Wu, J.; Zhang, Y.; Wang, F., Ultrafast charge transfer in atomically thin MoS2/WS2 heterostructures. *Nature Nanotechnology* **2014,** *9* (9), 682-686.
26. Li, T.; Li, M.; Lin, Y.; Cai, H.; Wu, Y.; Ding, H.; Zhao, S.; Pan, N.; Wang, X., Probing Exciton Complexes and Charge Distribution in Inkslab-Like WSe2 Homojunction. *ACS Nano* **2018,** *12* (5), 4959-4967.
27. Deilmann, T.; Thygesen, K. S., Interlayer Trions in the MoS2/WS2 van der Waals Heterostructure. *Nano Lett* **2018,** *18* (2), 1460-1465.
28. Ross, J. S.; Wu, S.; Yu, H.; Ghimire, N. J.; Jones, A. M.; Aivazian, G.; Yan, J.; Mandrus, D. G.; Xiao, D.; Yao, W.; Xu, X., Electrical control of neutral and charged excitons in a monolayer semiconductor. *Nat Commun* **2013,** *4*, 1474.
29. Jiang, C.; Xu, W.; Rasmita, A.; Huang, Z.; Li, K.; Xiong, Q.; Gao, W. B., Microsecond dark-exciton valley polarization memory in two-dimensional heterostructures. *Nat Commun* **2018,** *9* (1), 753.
30. Rivera, P.; Schaibley, J. R.; Jones, A. M.; Ross, J. S.; Wu, S.; Aivazian, G.; Klement, P.; Seyler, K.; Clark, G.; Ghimire, N. J.; Yan, J.; Mandrus, D. G.; Yao, W.; Xu, X., Observation of long-lived interlayer excitons in monolayer MoSe2-WSe2 heterostructures. *Nat Commun* **2015,** *6*, 6242.
31. Plechinger, G.; Nagler, P.; Arora, A.; Schmidt, R.; Chernikov, A.; del Águila, A. G.; Christianen, P. C. M.; Bratschitsch, R.; Schüller, C.; Korn, T., Trion fine structure and coupled spin–valley dynamics in monolayer tungsten disulfide. *Nature Communications* **2016,** *7*, 12715.
32. Zhang, X. X.; You, Y.; Zhao, S. Y.; Heinz, T. F., Experimental Evidence for Dark Excitons in Monolayer WSe_{2}. *Phys Rev Lett* **2015,** *115* (25), 257403.
33. Lee, J.; Wang, Z.; Xie, H.; Mak, K. F.; Shan, J., Valley magnetoelectricity in single-layer MoS2. *Nature Materials* **2017,** *16* (9), 887-891.
34. Zeng, H.; Dai, J.; Yao, W.; Xiao, D.; Cui, X., Valley polarization in MoS2 monolayers by optical pumping. *Nat Nanotechnol* **2012,** *7* (8), 490-3.
35. Aivazian, G.; Gong, Z.; Jones, A. M.; Chu, R.-L.; Yan, J.; Mandrus, D. G.; Zhang, C.; Cobden, D.; Yao, W.; Xu, X., Magnetic control of valley pseudospin in monolayer WSe2. *Nature Physics* **2015,** *11* (2), 148-152.
36. Zhang, X. X.; Cao, T.; Lu, Z.; Lin, Y. C.; Zhang, F.; Wang, Y.; Li, Z.; Hone, J. C.; Robinson, J. A.; Smirnov, D.; Louie, S. G.; Heinz, T. F., Magnetic brightening and control of dark excitons in monolayer WSe2. *Nat Nanotechnol* **2017,** *12* (9), 883-888.
37. Yu, H.; Wang, Y.; Tong, Q.; Xu, X.; Yao, W., Anomalous Light Cones and Valley Optical Selection Rules of Interlayer Excitons in Twisted Heterobilayers. *Phys Rev Lett* **2015,** *115* (18), 187002.
38. Seyler, K. L.; Rivera, P.; Yu, H.; Wilson, N. P.; Ray, E. L.; Mandrus, D. G.; Yan, J.; Yao, W.; Xu, X.,

Signatures of moire-trapped valley excitons in MoSe2/WSe2 heterobilayers. *Nature* **2019,** *567* (7746), 66-70.

39. Jones, A. M.; Yu, H.; Ross, J. S.; Klement, P.; Ghimire, N. J.; Yan, J.; Mandrus, D. G.; Yao, W.; Xu, X., Spin–layer locking effects in optical orientation of exciton spin in bilayer WSe2. *Nature Physics* **2014,** *10* (2), 130-134.

40. Lee, P. A.; Ramakrishnan, T. V., Disordered electronic systems. *Reviews of Modern Physics* **1985,** *57* (2), 287-337.

41. Srivastava, A.; Sidler, M.; Allain, A. V.; Lembke, D. S.; Kis, A.; Imamoğlu, A., Valley Zeeman effect in elementary optical excitations of monolayer WSe2. *Nature Physics* **2015,** *11* (2), 141-147.

42. Li, Y.; Ludwig, J.; Low, T.; Chernikov, A.; Cui, X.; Arefe, G.; Kim, Y. D.; van der Zande, A. M.; Rigosi, A.; Hill, H. M.; Kim, S. H.; Hone, J.; Li, Z.; Smirnov, D.; Heinz, T. F., Valley Splitting and Polarization by the Zeeman Effect in Monolayer ${\mathrm{MoSe}}_{2}$. *Physical Review Letters* **2014,** *113* (26), 266804.

43. Hsu, W.-T.; Zhao, Z.-A.; Li, L.-J.; Chen, C.-H.; Chiu, M.-H.; Chang, P.-S.; Chou, Y.-C.; Chang, W.-H., Second Harmonic Generation from Artificially Stacked Transition Metal Dichalcogenide Twisted Bilayers. *ACS Nano* **2014,** *8* (3), 2951-2958.

## Acknowledgements

We acknowledge the financial support from the Ministry of Science and Technology of China (Grant 2016YFA0200600), the National Natural Science Foundation of China (Grants 11474260, 21573207, 11504359, 11504364 and 11804332), the Anhui Initiative in Quantum Information Technologies (Grant AHY090200), and the Fundamental Research Funds for the Central Universities (Grant WK2060190084). T. Y. was supported by the Netherland Organization for Scientific Research (NWO). The authors also thank Prof. Jin Zhao and Mr. Xiang Jiang for their helpful discussion.

## Author contributions

T. S. Li, N. Pan, and X. P. Wang designed and conducted this research. T. S. Li prepared all of the samples, and performed the experiments with the assistance of X. F. Cui, K. X. Zhang, J. Y. Liu, and Q. S. Meng. T. Yu and T. S. Li established the theoretical model in this work. T. S. Li, T. Yu, N. Pan and X. P. Wang analyzed the experiment data and co-wrote the paper. All authors contributed to the discussions.

## Figures

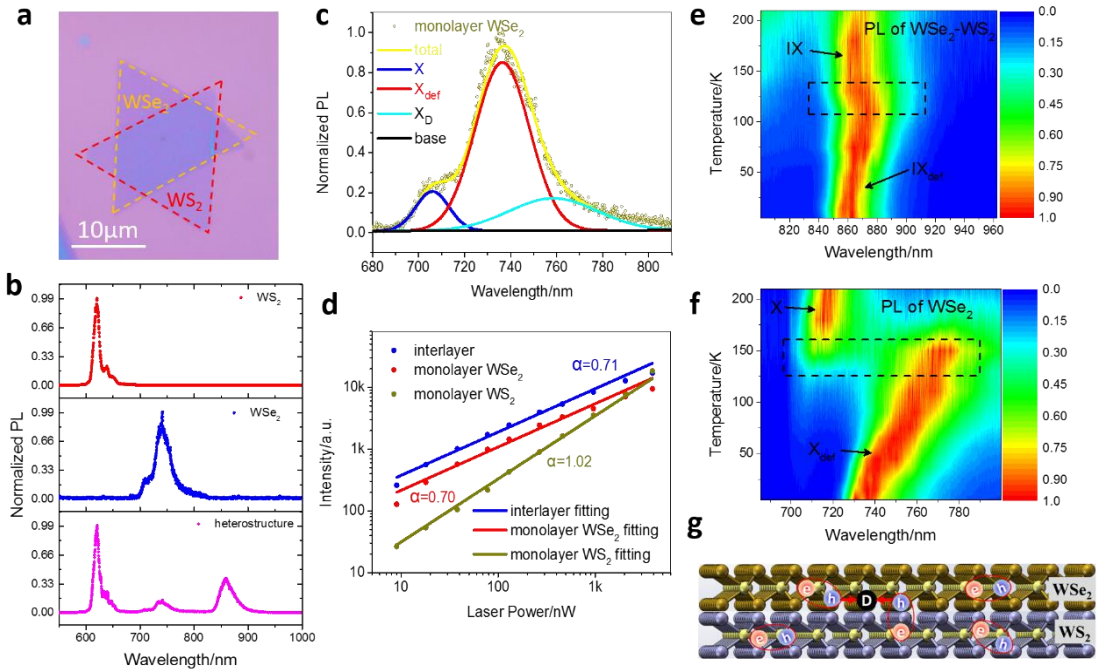

Fig. 1 PL properties and band alignment of AB-stacked WSe$_2$-WS$_2$ heterostructure. a, The optical microscope image of AB-stacked WSe$_2$-WS$_2$ heterostructure. b, Normalized PL properties of monolayer WS$_2$ zone (red curve), monolayer WSe$_2$ zone (blue curve) and heterostructure zone (purple curve) in the sample, respectively. c, Fitting result of the PL of monolayer WSe$_2$ shown in Fig. 1b. d, Excitation power dependent PL properties, the colorful dots and lines correspond to experiment data and fitting results, respectively. e and f, Normalized PL spectrum corresponding to the heterostructure and monolayer WSe$_2$ zone as a function of temperature, respectively. g, schematic of exciton configuration in real space of AB-stacked WSe$_2$-WS$_2$ heterostructure.

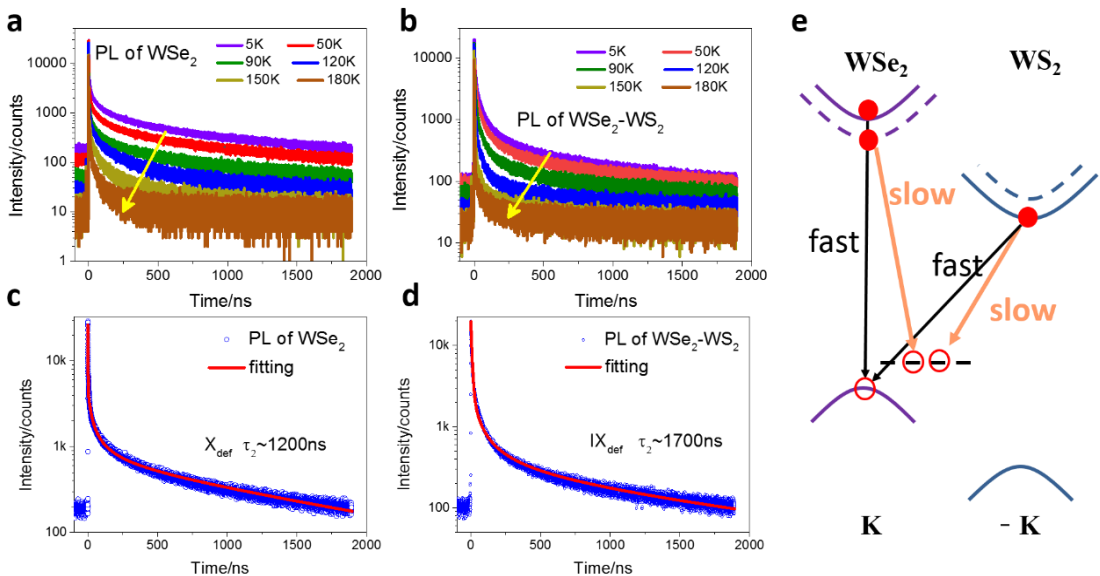

Fig. 2 Time-resolved PL and schematic of transition process of defect related excitons in AB-stacked WSe$_2$-WS$_2$ heterostructure. a and b, Time-resolved PL of monolayer WSe$_2$ (a) and WSe$_2$-WS$_2$ heterobilayer (b) at different temperature. c and d, Experimental (blue dots) and fitting (red lines) time-resolved PL of monolayer WSe$_2$ (c) and WSe$_2$-WS$_2$ heterobilayer (d) at temperature 5K. e,

Diagram of transition process of defect related excitons in AB stacking WSe$_2$-WS$_2$ heterostructure. The excitation wavelength of pulse laser is at 670nm with repetition 500 KHz.

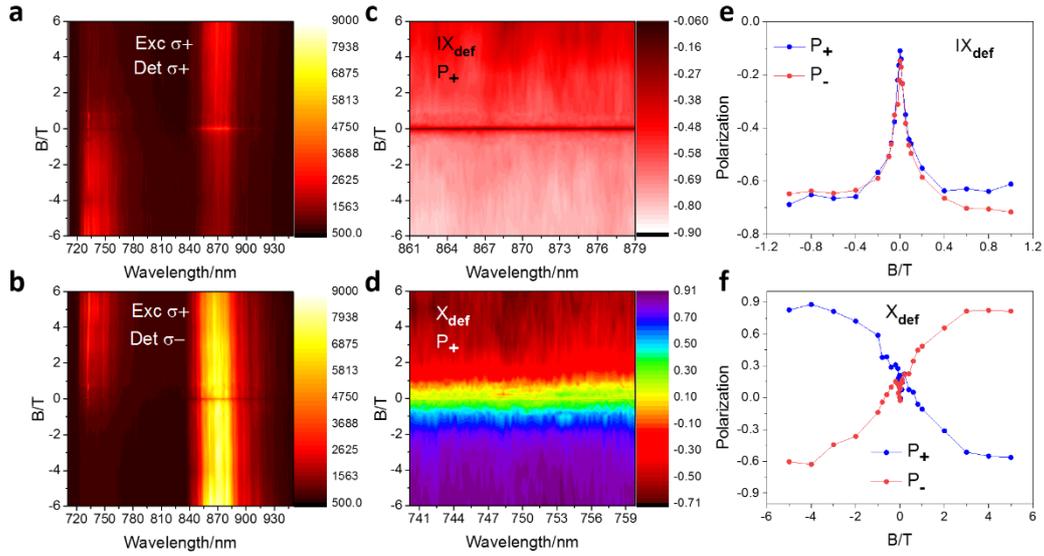

Fig. 3 Out-of-plane magnetic-field tunable PL intensity and polarization from IX$_{def}$ and X$_{def}$ in the AB-stacked WSe$_2$-WS$_2$ heterostructure. a and b, σ+ (a) and σ- (b) detected polarization-resolve PL spectrum as a function of out-of-plane magnetic field with σ+ excitation at 670 nm wavelength. c and d, Out-of-plane magnetic field tuning polarization degree of IX$_{def}$ (c) and X$_{def}$ (d) with σ+ excitation, respectively. e and f, The averaged valley polarization of IX$_{def}$ (e extracted from Fig. 3c and S3c) and d and X$_{def}$ (f extracted from Fig. 3d and S3d) as a function of B$_z$, respectively.

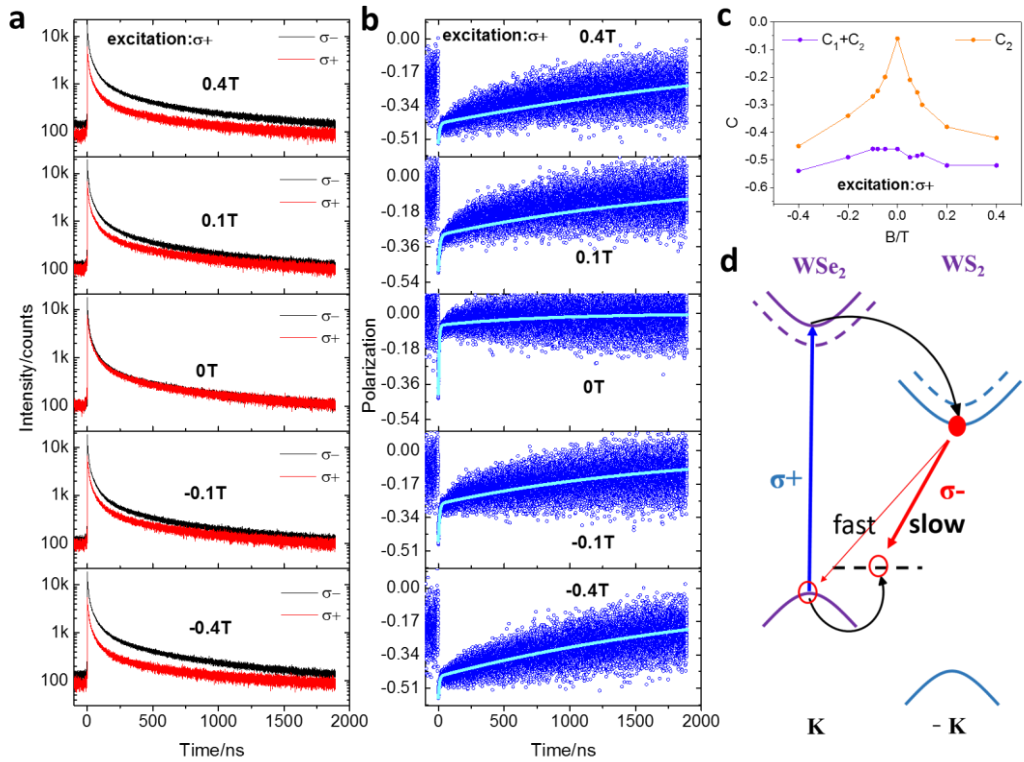

Fig. 4 The magnetic-field tunable time-resolved PL and valley polarization and schematics of transition dynamics of IX$_{def}$. a, Time- and polarization-resolved PL of WSe$_2$-WS$_2$ heterobilayer with σ+ excitation at selected out-of-plane magnetic fields. b, Time-resolved PL polarization calculated

from Fig. 4a at selected out-of-plane magnetic fields. The blue dots denote the experimental data and the light blue lines represent the fitting results. c, Fitting results of initial polarization of fast ($C_1+C_2$) and slow ($C_2$) process under σ+ excitation respectively. d, Schematics of transition dynamics of $IX_{def}$.

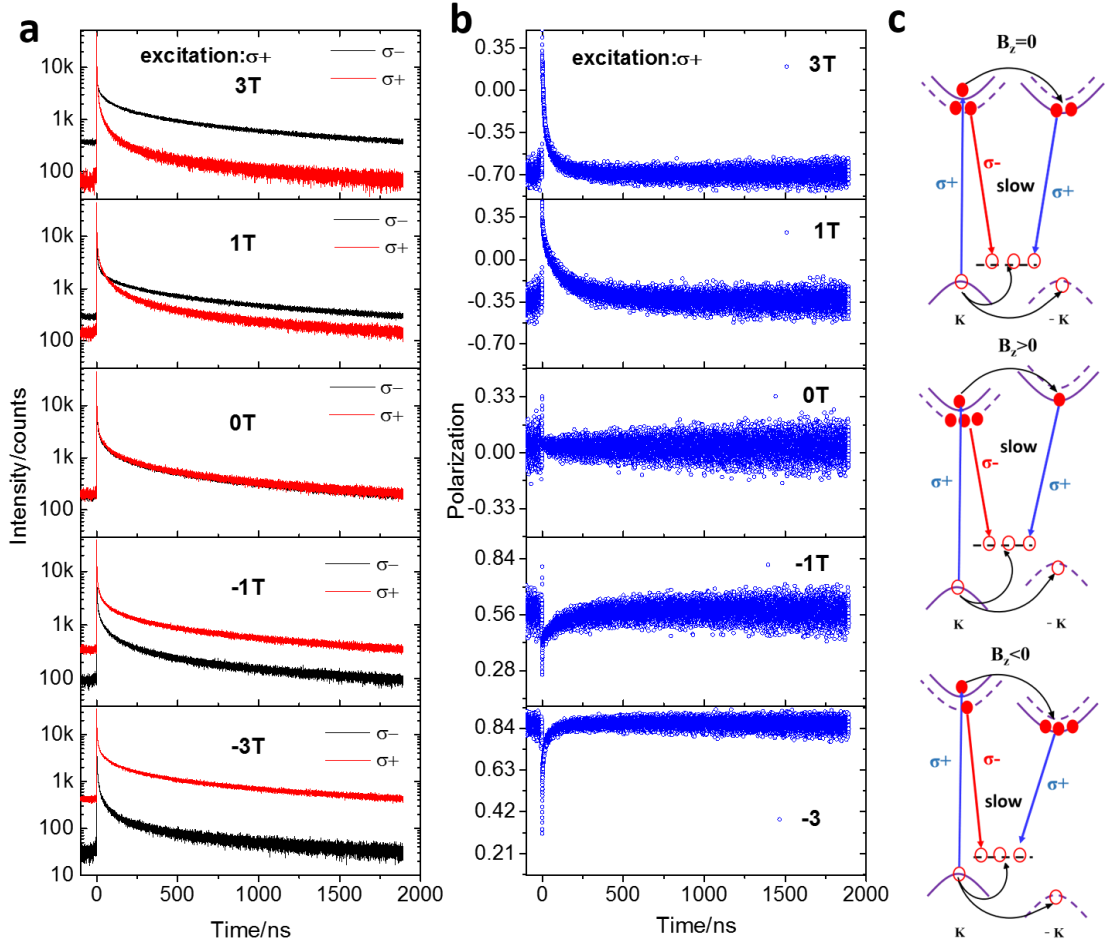

Fig. 5 Magnetic-field tunable time-resolved PL and valley polarization dynamics and schematic of valley dynamics of $X_{def}$. a, Time- and polarization-resolved PL of monolayer WSe$_2$ with σ+ excitation at selected out-of-plane magnetic fields. b, Time-resolved PL polarization of monolayer WSe$_2$ with σ+ excitation at selected out-of-plane magnetic fields. c, Schematic of valley dynamics of $X_{def}$ under zero, positive and negative out-of-plane magnetic fields, respectively.